# Experimental observations of temperature effects in the near-field regime of the Casimir-Polder interaction

**T PASSERAT DE SILANS[1,2], A LALIOTIS[1], I MAURIN[1], M-P GORZA[1], P CHAVES DE SOUZA SEGUNDO[3], M DUCLOY[1] and D BLOCH[1]**

[1] *Laboratoire de Physique des Lasers , Université Paris 13- Sorbonne Paris-Cité, CNRS UMR 7538, 99, av. JB Clément F-93430 Villetaneuse*
[2] *Laboratório de Superfícies, D. F., Universidade Federal da Paraíba, Cx. Postal 5086, 58051-900, João Pessoa, PB, Brazil*
3 *Unidade Academica de Educação.CES Campus Cuité, Universidade Federal de Campina Grande, Cuité, PB, Brazil*

email: athanasios.laliotis@univ-paris13.fr

*We investigate the temperature dependence of the Casimir-Polder interaction in the electrostatic limit. This unusual phenomenon relies on the coupling between a virtual atomic transition and a thermal excitation of surface polariton modes. We first focus on the scenario where a $Cs(8P_{3/2})$ atom is next to a $CaF_2$ or a $BaF_2$ surface. Our theoretical predictions show a strong temperature dependence of the van der Waals coefficient at experimentally accessible conditions. A series of spectroscopic measurements performed in a specially designed Cs vapour cell containing a $CaF_2$ tube is presented. Our results illustrate the sensitivity of atom surface-interaction experiments to the quality and chemical stability of the surface material and emphasize the need of using more durable materials, such as sapphire. We finally discuss selective reflection experiments on $Cs(7D_{3/2})$ in an all-sapphire cell that clearly demonstrate a temperature dependent van der Waals coefficient.*

1. INTRODUCTION

Interactions between neutral but polarisable objects are ubiquitous and fundamental for our understanding of the physical world. A typical example is the long range atom-surface force, most commonly known as the Casimir-Polder interaction [1]. In the near field it is most commonly described as the interaction between the fluctuating dipole and its image. While in the far field, it becomes easier to portray it as a distance-dependent Lamb shift, due to the modification of vacuum or thermal fluctuations by the reflective boundary [2].

The temperature dependence of the Casimir-Polder interaction is an important consideration when comparing theory and experiment, and temperature corrections are fundamental for any precision measurement involving atoms close to surfaces [3,4]. On a more practical side, atom surface interactions at non zero temperature can be of relevance for atom or molecule chips [5-9] and for miniaturized frequency references or clocks [10-11].

While the effects of thermal fluctuations on the Lamb shift were demonstrated in a high precision experiment in the early eighties [12], the Casimir-Polder interaction was observed several years later [13] and its temperature dependence was more elusive. The only measurement demonstrating such effects was made at distances on the order of 5-10 μm where the strength of the interaction is extremely small. Critical to experimental success [14] was an enhancement due to an out of equilibrium effect [15], where the surface is held at a higher temperature with respect to its surrounding environment.

Taking a look much closer to the surface, typically around 100 nm, the shift of an atomic level takes the familiar form $-C_3/z^3$, where $C_3$ is the van der Waals coefficient. The classical picture of an atom interacting with its image is less accommodating for thermal effects, since the reflected field of a dipole antenna does not fundamentally depend on temperature [16,17]. This intuition is supported by quantum electrodynamics (QED) calculations for an atom or molecule in front of a perfect or dispersion-less conductor [18].

Material dispersion and especially surface polaritons are, however, of great importance in the near field. In this regime thermal emission is dominated by surface excitations and becomes almost monochromatic in contrast to the well-known broadband black body spectrum [19,20]. Similarly the van der Waals interaction is dominated by the resonant coupling between atom dipole transitions and surface polaritons. This is usually relevant for excited state atoms whose couplings in the mid infrared range coincide with surface excitations of common dielectrics. It was demonstrated that the coupling between a virtual atomic emission and a surface mode excitation can lead to extraordinary effects such as atom-surface repulsion or even giant attraction [21,22]. The inverse effect requires a thermal excitation of the polariton modes that can now be coupled to an atomic absorption and leads to a temperature dependent van der Waals interaction [2,21,23,24]. This exotic behaviour is only accessible when the thermal energy is comparable to that of the atomic transition (i.e. $\hbar\omega_o=k_BT$). For room temperatures this dictates a transition wavelength on the order of 50 μm.

The above reasoning essentially excludes ground state atoms and in general puts heavy constraints on the appropriate atom surface system. An extensive study of materials [25] suggests that the $Cs(8P_{3/2})$ dipole couplings $8P_{3/2} \rightarrow 7D_{3/2}$ and $8P_{3/2} \rightarrow 7D_{5/2}$ at 39μm and 36 μm respectively (Fig.1) can couple resonantly to $CaF_2$ and $BaF_2$ surface polaritons at 24 μm and 35 μm respectively. The dielectric properties of these materials were experimentally measured as a function of temperature [26]. Here we start by detailing the theoretical predictions of the temperature dependence of the van der Waals coefficient for $Cs(8P_{3/2})$ against a $CaF_2$, $BaF_2$ and a sapphire surface. We then proceed to describe the corresponding experiments which were inconclusive due to surface quality deterioration. We finally give a flavour of a successful series of experiments on $Cs(7D_{3/2})$ at a sapphire interface which will be reported in detail elsewhere.

## 2. THEORETICAL PREDICTIONS

In the near field, the atom surface interaction induces a free energy shift which is given by $-C_3/z^3$. The van der Waals coefficient depends on the surface, as well as the atom and its given energy state |i>. To calculate $C_3$ one needs to sum the contribution of all allowed dipole couplings |i> $\rightarrow$ |j> which for a perfect reflector is proportional to the square of the electric dipole moment matrix element $<i|D|j>$. In the case of a real surface this has to be multiplied by an image coefficient $r_{ij}$, which depends on the dielectric properties of the surface and on. Ignoring the dispersive response of the material, the image coefficient is simply given by $r = \frac{\epsilon-1}{\epsilon+1}$ , independent of ambient temperature, and consistent with the classical picture of a dipole interacting with its own image in front of a surface with finite reflectivity. A complete QED description [16] is required to account for surface dispersion. A few theoretical studies have dealt with this problem in the past [23-25] showing that there are two contributions to the image coefficient. The first resembles a distance dependent Lamb shift due to vacuum and thermal fluctuations covering the entire frequency spectrum. The second is reminiscent of the interaction between a classical oscillating dipole and its own reflected field [27]. For convenience the former, dispersive, contribution will be henceforth referred to as non-resonant, $r_{nr}$, whereas the latter as resonant, $r_{res}$.

For an upward coupling |i>→|j> the resonant term takes the form:

$$r_{res} = -2Re[S(\omega_o)]\, n(\omega_o, T) \qquad (1)$$

where $\omega_o = \omega_j - \omega_i > 0$ is the transition frequency and the factor $S(\omega_o) = \frac{\epsilon(\omega_o)-1}{\epsilon(\omega_o)+1}$ is the complex surface response from which the polariton modes are defined and $n(\omega_o, T) = \frac{1}{e^{\frac{\hbar\omega_o}{kT}}-1}$ is the mean occupation number of each mode by photons. The term in eqn. (1) owes its existence to temperature. In the case of a downward transition, $\omega_o < 0$, the resonant term has a similar temperature dependence given by $n(\omega_o, T) = n(|\omega_o|) + 1$ but survives even at zero temperatures due to spontaneous emission [23].

In reality an atomic level |i> has numerous couplings and isolating the resonant term of a given transition is extremely difficult. In Fig.2 we show the real part of the surface response as a function of frequency for calcium fluoride, barium fluoride and sapphire with surface resonances at 24 μm, 35 μm, and 12 μm respectively. In Table 1 we show the temperature dependence of the most dominant dipole couplings, starting from the 8P$_{3/2}$ level, for each of these three materials as calculated using the QED theory described in [23] and data for the dielectric constants given in [26]. As can be verified in the Table 1, the temperature dependence of the C$_3$ coefficient is predominantly due to dipole couplings that are close to the respective polariton resonances. In the case of BaF$_2$ and CaF$_2$ these are the 8P$_{3/2}$→7D$_{5/2}$ and 8P$_{3/2}$→7D$_{3/2}$ couplings. In the case of sapphire these couplings fall clearly out of the range of the surface polariton and their temperature dependence is modest. It is worth noticing that the 8P$_{3/2}$→9S$_{1/2}$ coupling at 8,9 μm, which is closer to the sapphire surface resonance, practically cancels any temperature dependence induced by the dominant coupling 36μm. A close examination of Table 1 is very instructive. It reveals the complexity of the C$_3$ dependence on temperature and illustrates the importance of Casimir-Polder type experiments for our understanding of the electromagnetic properties of materials and surfaces. It is also worth mentioning that the temperature dependence of the dielectric constant itself [26] has been here neglected when calculating the values in Table 1. In Fig.2 the dotted lines show the surface response when the surface is at T=770 K. The effects are dramatic for the polariton resonances but not for the rest of the spectrum. Our predictions for C$_3$ are almost unaffected in the case of sapphire and CaF$_2$, but they need to be seriously revised for BaF$_2$ [26].

### 3.    SELECTIVE REFLECTION EXPERIMENTS ON THE 6S→8P LINE

Frequency Modulated (FM) selective reflection at the interface between a window and vapour is one of the few linear spectroscopic methods providing a signal with sub-Doppler resolution [10,28-30]. Extensive theoretical and experimental studies have shown that, in its simplest form, it is sensitive to atoms typically within a distance of λ/2π from the surface of the window. This unique characteristic makes it ideal to probe atom surface interactions in the near field regime [30-32]. The van der Waals coefficient is measured by comparing the experimentally obtained spectra to a library of theoretical curves [22]. Using a fitting process we extract C$_3$, the transition linewidth Γ, and the collisional shift δ. The above well-established method has been repeatedly used in the past to measure the van der Waals coefficient [30-37]. Note that in most cases the shift of a given transition is governed solely by the van der Waals interaction of the high lying state.

The experiments that are reported here are performed on the third resonance line of Cs at 388 nm [37-38]. Initially we used an extended cavity laser system with a diode emitting at the UV spectral range. Unfortunately the power output was limited to 1-2 mW and the beam quality was low. Eventually this source was replaced by an amplified, frequency-doubled 780 nm laser diode with a final output power of about 100mW at UV wavelengths. This frequency source had the additional advantage of allowing us to scan both the $6S_{1/2} \rightarrow 8P_{1/2}$ line at 387.6 nm and the $6S_{1/2} \rightarrow 8P_{3/2}$ line at 388.8 nm. Both sources were frequency modulated by applying a voltage on the piezoelectric element attached on the grating of the extended cavity laser. The frequency doubled source was modulated by double passing the beam through an acousto-optic modulator. A saturated absorption was performed in a slightly heated (~80$^o$C) sapphire vapour cell. Additionally we used a stable Fabry-Perot cavity with a free spectral range of 83 MHz as a frequency marker. The above auxiliary experiments allowed us to determine the absolute frequency of the laser throughout our scans with an accuracy of a few MHz.

In Fig.3 we show the vapour cell in which selective reflection measurements were performed. The design was identical to the one described in [35] but this time no impurities were present. The fabrication of a vapour cell with $CaF_2$ windows was technically challenging due to the different thermal expansion coefficients between the window and the main body of the cell which is made out of sapphire. The cell is T-shaped with one window made of sapphire and the other one made of YAG. A hole is drilled on the main body of the cell, onto which the sidearm is attached. A $CaF_2$ tube is inserted inside the cell, almost in contact with the YAG window. The small gap between the tube and the window is about 100 μm. This part of the cell is kept at low temperatures to ensure that Cs density within the gap stays negligible and does not affect the selective reflection which is performed on the $CaF_2$ interface. The temperature of the cell is controlled by independent ovens. The first and second ovens control the temperature of the sapphire window, and the $CaF_2$ tube respectively. The third controls the sidearm (reservoir) temperature and therefore the Cs density inside the cell. To avoid Cs condensation on the tube or the window the upper part of the cell is always kept at slightly higher temperatures than the sidearm. Repeated attempts to fabricate a similar cell with a $BaF_2$ tube were unsuccessful due to the fragility and chemical instability of the material at high temperatures [39].

Typically we record selective reflection spectra on the $CaF_2$ tube and the sapphire window simultaneously. Our well established experimental protocol is as follows: First, for a given window temperature we vary the Cs vapour pressure, therefore changing the transition linewidth due to pressure broadening and the collisional shift. The same process is repeated for both the F=3 → F'=2,3,4 and the F=4 → F'=3,4,5 manifold of the $6S_{1/2} \rightarrow 8P_{3/2}$ transition whose spectra are fundamentally different due to the different relative weight of the hyperfine transitions. Unfortunately it became very soon evident that the quality of the $CaF_2$ tube started to deteriorate after use at high temperatures in the presence of chemically aggressive Cs vapour. The most striking effect was the fact that the material became porous and Cs in gas phase infiltrated the tube. This meant that a parasitic Doppler shaped absorption due to Cs atoms inside the tube was always superimposed with the selective reflection spectrum.

We nonetheless finished an extensive series of measurements. In Fig.4 we show two simultaneously recorded selective reflection spectra on the $CaF_2$ and sapphire interfaces. Initially we attempted to fit the spectra by imposing the same parameters for the vapour linewidth Γ (including pressure broadening) and for the collisional shift δ, as compared with the saturated absorption reference, which should remain a fraction of $\Gamma$. This is a reasonable restriction under the assumption that the vapour pressure within the cell is only controlled by the Cs reservoir in the sidearm of the cell. However, in our case the porous $CaF_2$ tube also acts as an independent reservoir making it impossible to assign a uniform value to $\Gamma$ and $\delta$ throughout the cell. We therefore fitted each spectrum independently and this gave satisfactory results as can be verified in Fig.4. Given the state of the $CaF_2$ tube it was rather surprising that all of measurements and fits allowed us to consistently extract the values for the $C_3$ coefficient that are shown in Fig.5 as a function of the window temperature. The experimental results are at odds with theoretical predictions, also shown in the same figure.

Finally we attempted to fit our experimental spectra using a potential of the form $-C_2/z^2$ to describe the interaction between the atom and the surface. This potential could result if the surface is electrically charged or somehow contaminated with Cs adsorbents [4,39,40]. For this purpose a new library of curves was produced and the fitting process was repeated using $C_2$, $\Gamma$ (linewidth) and $\delta$ (collisional shift) as independent parameters. The quality of the fits was satisfactory, however, the values of $\Gamma$ and $\delta$ required in order to produce those fits were unreasonable and inconsistent with previous spectroscopic measurements on the $6S_{1/2} \rightarrow 8P_{3/2}$ line [37]. In Fig. 6 we can see the collisional shift, as extracted from the fits of experimental spectra using both $-C_3/z^3$ (solid points) and $-C_2/z^2$ (open points). Data are presented for both sapphire (circles) and $CaF_2$ (squares) window. It is clear that even at low pressures one needs to impose a large positive collisional shift when fitting with a $-C_2/z^2$ potential, something that is artificial in normal conditions. Moreover, the previously reported collisional shift was on the order of -60 MHz/Torr [37], a measurement that is only supported when fitting with a $-C_3/z^3$ potential.

The above arguments lead to the conclusion that our measurements cannot be explained with a $-C_2/z^2$ potential and that we can have relative confidence on the values extracted for the van der Waals coefficient. It seems, however, that the chemical nature of the surface was significantly altered and this dramatically modifies the polariton resonances relatively to what we have supposed for our theoretical calculations. Note in particular that an ideal roughness is assumed when converting the bulk properties of the material from measured values of the dielectric constant into surface polariton modes.

## 4. SELECTIVE REFLECTION EXPERIMENTS ON THE 6P→7D LINE

The fragility of $CaF_2$ has forced us to turn our attention to different materials, such as sapphire which has been used for this purpose in the past and has been remarkably stable in the presence of alkali vapours [42,43]. Moreover, technology allows the fabrication of all-sapphire vapour cells that can withstand very high temperatures. For our selective reflection experiments that are naturally sensitive to the quality of the windows we have acquired specially designed all-sapphire vapour cell with a sidearm that can be heated up to almost 1000°C made in the group of D. Sarkisyan. The main window is super-polished with an average surface roughness of about 0,3 nm. This cell replaced a similar but much older one with windows of unknown quality that was initially used for our measurements.

As can be verified in Table 1 the van der Waals coefficient for Cs(8P) near sapphire displays no temperature dependence [26]. We need therefore to find another atomic level for which a dominant dipole coupling falls closer to the sapphire polariton at 12 μm. An obvious candidate is the Cs($7P_{1/2}$) whose coupling with $6D_{3/2}$ is at 12.15 μm [21]. Probing the weak $6S_{1/2} \rightarrow 7P_{1/2}$ transition at 459 nm is, however, difficult due to the lack of low noise laser sources at this wavelength [34]. Moreover this coupling is so close to the sapphire polariton that it becomes sensitive to the surface quality and therefore challenging to ascertain a theoretical prediction.

Instead we chose Cs($7D_{3/2}$) as our first candidate. This level has an upward coupling, $7D_{3/2} \rightarrow 5F_{5/2}$ at 10.8μm (Fig.1) while the couplings at 36μm and 39μm fall far away from the sapphire resonance (Fig.2). The relatively small transition wavelength suggests that temperature needs to be substantially increased, before a change of the van der Waals coefficient can be observed. For our experiments the atoms are first pumped with a high power laser to the Cs($6P_{1/2}$) level and from there selective reflection is performed with a red laser on the $6P_{1/2} \rightarrow 7D_{3/2}$ transition at 672nm. This time we take measurements for window temperatures as high as 1000 K, with a significant increase of $C_3$ from ~ 50 kHz μm$^3$ at 500 K to ~ 80 kHz μm$^3$ at 1000 K. The results of our experiments in both all-sapphire cells verify our theoretical predictions. A typical spectrum obtained for the F=4 $\rightarrow$ F'=3,4,5 manifold of the $6P_{1/2} \rightarrow 7D_{3/2}$ transition is shown in Fig.7. The quality of the fit is impressive allowing us to measure the van der Waals coefficient with a much better than 15% from a single spectrum. More details on these experiments will be given in an upcoming publication [44].

## 5. CONCLUSIONS

We have given an overview of selective reflection experiments aimed at measuring the temperature dependence of the Casimir-Polder interaction between an atom and a surface in the near-field regime. Initial measurements were performed in a specially designed cell containing a $CaF_2$ tube on the $6S_{1/2} \rightarrow 8P_{3/2}$ transition of Cs. Even though theoretical estimates predicted a strong variation of $C_3$ with temperature, this was not observed. The reason behind this surprising fact lies on the deterioration of the $CaF_2$ tube, which makes impossible to predict the polariton modes on which the temperature dependence of the van der Waals coefficient critically relies. For this reason we did not further pursue the fabrication of a $BaF_2$ cell even though the predicted thermal effects are huge. Instead we focused on an all sapphire cell. Our measurements on the $6P_{1/2} \rightarrow 7D_{3/2}$ line of Cs show a dependence of $C_3$ with temperature which follows our QED calculations. It is clear from our theoretical predictions that the temperature dependence of the van der Waals interaction is very specific to the atom-surface system in question. In the future one could take advantage of this phenomenon in order to tune or even eliminate atom surface interactions at a specific temperature.

ACKGNOWLEDGEMENTS

We would like to acknowledge financial support from CAPES-COFECUB (Ph 740-12) and useful discussions with J R Rios Leite, H Failache and M Chevrollier. It has also been a pleasure to collaborate with D. Sarkisyan who fabricated both the $CaF_2$ and the all-sapphire cell for our experiments. We finally would like to thank G Pichler for kindly offering us his all-sapphire cell in which initial experiments on the Cs($7D_{3/2}$) were performed.

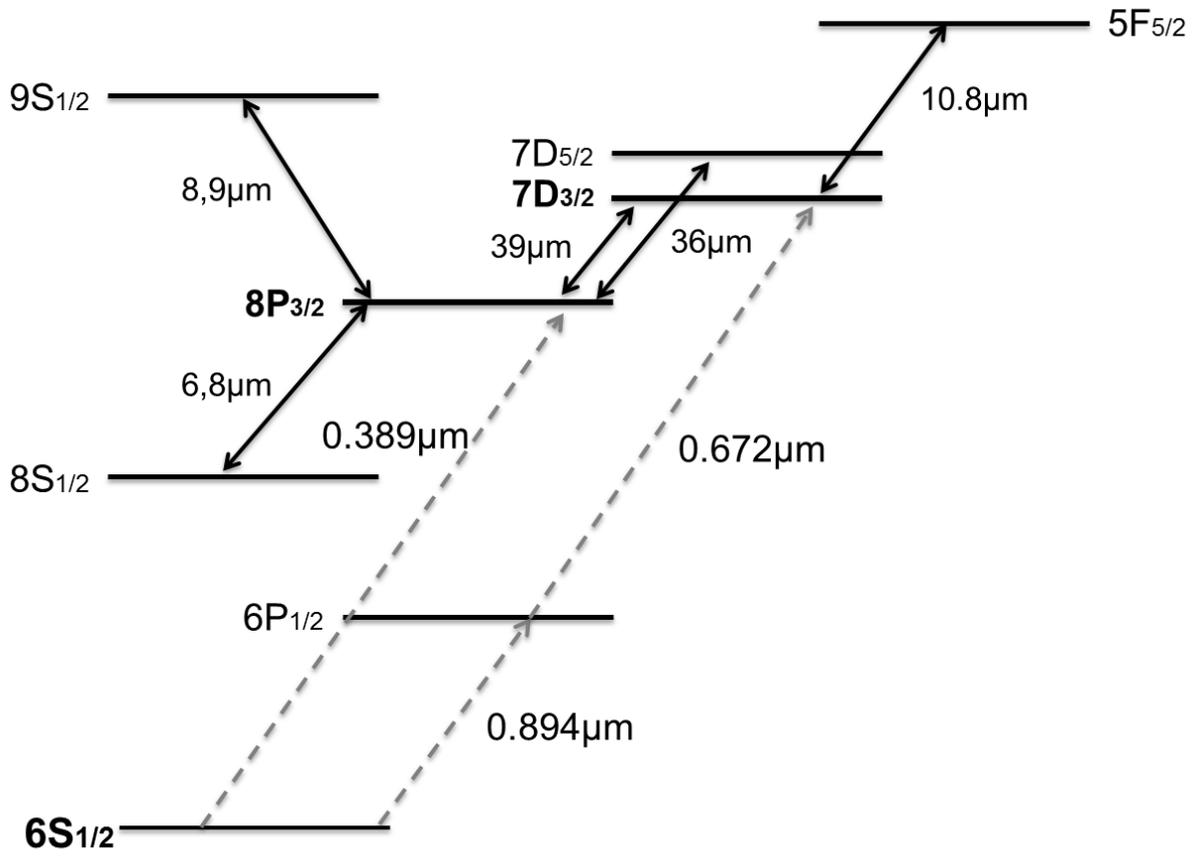

**Figure 1.** Schematic diagram of Cs energy levels relevant to our experiments.

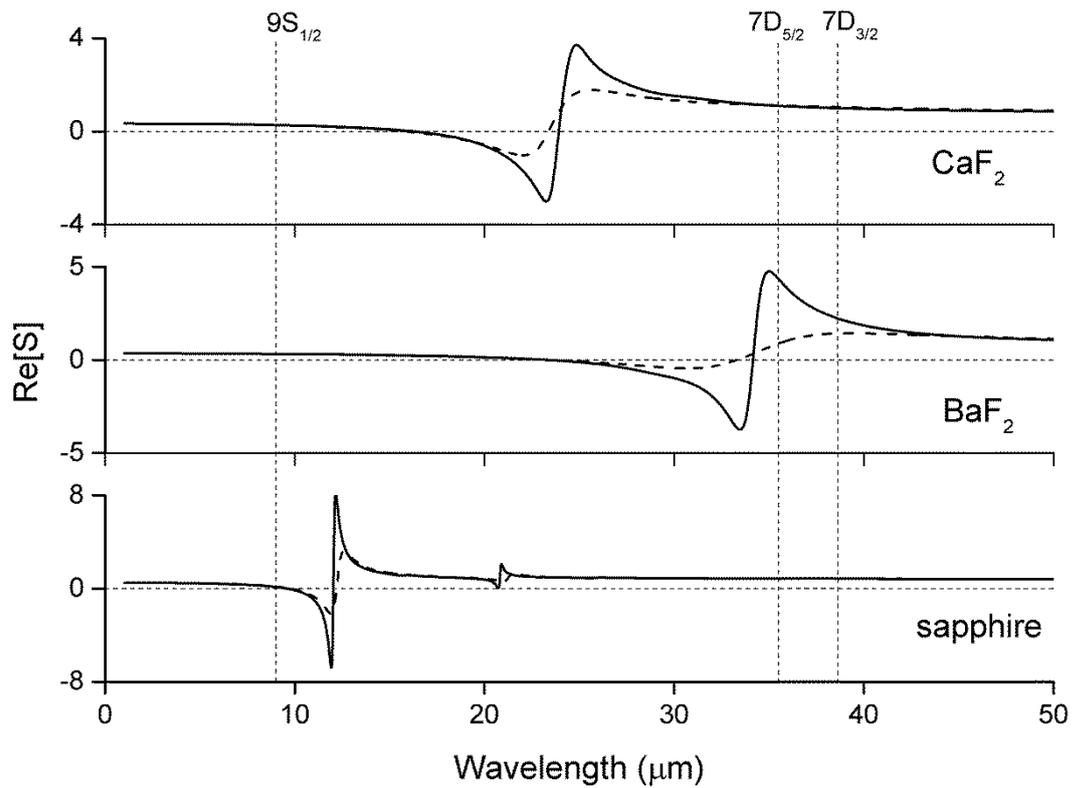

**Figure 2.** The real part of the image coefficient for $CaF_2$, $BaF_2$ and sapphire as a function of wavelength [26]. The solid curve represents measurements at room temperature whereas the dashed curve represents measurements made at T=770 K. The position of the dominant dipole couplings is shown as a vertical dashed line. The dominant coupling $8P_{3/2} \rightarrow 7D_{5/2}$ at 36,09 μm is very close to the $BaF_2$ polariton at 35 μm, on the wings of the $CaF_2$ polariton at 24μm and very far away from the sapphire polariton at 12 μm.

**Table 1.** Contribution of each dipole coupling to the $C_3$ van der Waals coefficient between a Cs($8P_{3/2}$) and (a) CaF$_2$ (b) BaF$_2$ and (c) Sapphire for different temperatures. $C_3$ is measured in kHz µm$^3$ and the temperature in Kelvin. The negative sign denotes a downward transition. The $C_3$ value, given by the sum of each individual contribution, is also shown at the end of each table. The contributions to $C_3$ were calculated using measurements of the dielectric constant at room temperature [26].

(a) CaF2

| Cs($8P_{3/2}$) | λ (µm) | C3 (perfect reflector) | C3 (T=0) | C3 (T=200) | C3 (T=400) | C3 (T=600) | C3 (T=800) | C3 (T=1000) |
|---|---|---|---|---|---|---|---|---|
| $8S_{1/2}$ | -6,781 | 12,07 | 2,17 | 1,99 | 1,27 | 0,38 | -0,56 | -1,52 |
| $7D_{3/2}$ | 39,05 | 5,32 | 3,11 | 2,92 | 1,93 | 0,69 | -0,632 | -1,99 |
| $7D_{5/2}$ | 36,095 | 37,79 | 21,8 | 19,98 | 11,19 | 0,19 | -11,5 | -23,48 |
| $9S_{1/2}$ | 8,937 | 11,63 | 5,19 | 5,4415 | 6,41 | 7,61 | 8,88 | 10,19 |
| $8D_{5/2}$ | 4,923 | 3,7 | 1,51 | 1,5537 | 1,71 | 1,9 | 2,09 | 2,3 |
| Total | | 73,71 | 34,94 | 33,045 | 23,73 | 12,047 | -0,3677 | -13,0925 |

(b) BaF2

| Cs($8P_{3/2}$) | λ (µm) | C3 (perfect reflector) | C3 (T=0) | C3 (T=200) | C3 (T=400) | C3 (T=600) | C3 (T=800) | C3 (T=1000) |
|---|---|---|---|---|---|---|---|---|
| $8S_{1/2}$ | -6,781 | 12,07 | 3,14 | 2,86 | 2,09 | 1,23 | 0,34 | -0,56 |
| $7D_{3/2}$ | 39,05 | 5,32 | 3,01 | 0,71 | -5,84 | -13,19 | -20,77 | -28,43 |
| $7D_{5/2}$ | 36,095 | 37,79 | 21,07 | -11,44 | -105,46 | -211,34 | -320,54 | -431,09 |
| $9S_{1/2}$ | 8,937 | 11,63 | 5,13 | 5,49 | 6,49 | 7,62 | 8,79 | 9,97 |
| $8D_{5/2}$ | 4,923 | 3,7 | 1,52 | 1,58 | 1,75 | 1,94 | 2,13 | 2,33 |
| Total | | 73,71 | 35,06 | 0,42 | -99,69 | -212,4 | -328,64 | -446,329 |

(c) SAPPHIRE

| Cs($8P_{3/2}$) | λ (µm) | C3 (perfect reflector) | C3 (T=0) | C3 (T=200) | C3 (T=400) | C3 (T=600) | C3 (T=800) | C3 (T=1000) |
|---|---|---|---|---|---|---|---|---|
| $8S_{1/2}$ | -6,781 | 12,07 | 1,55 | 1,52 | 1,19 | 0,54 | -0,26 | -1,13 |
| $7D_{3/2}$ | 39,05 | 5,32 | 3,9 | 3,92 | 3,87 | 3,75 | 3,6 | 3,44 |
| $7D_{5/2}$ | 36,095 | 37,79 | 27,57 | 27,65 | 27,2 | 26,24 | 25,03 | 23,71 |
| $9S_{1/2}$ | 8,937 | 11,63 | 7,32 | 7,37 | 7,98 | 9,21 | 10,73 | 12,4 |
| $8D_{5/2}$ | 4,923 | 3,7 | 2,17 | 2,18 | 2,24 | 2,36 | 2,51 | 2,67 |
| Total | | 73,71 | 44,14 | 44,27 | 44,13 | 43,78 | 43,32 | 42,84 |

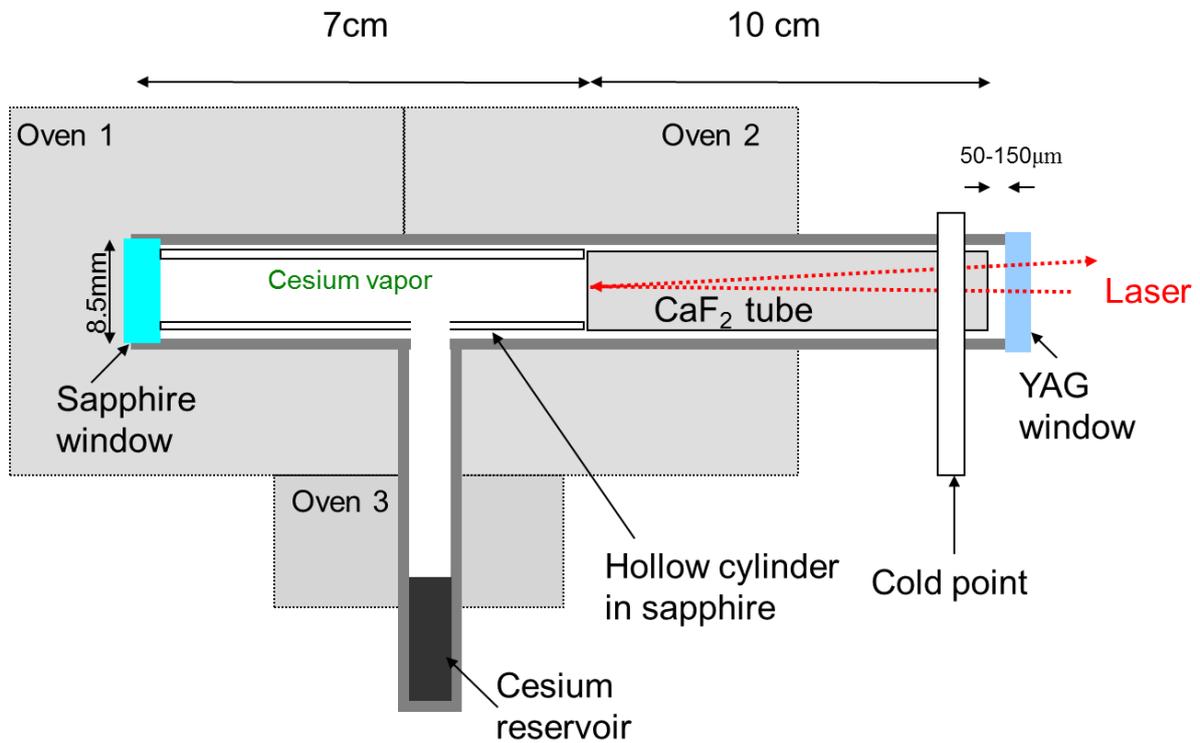

**Figure 3.** Schematic diagram of the Cs vapour cell, containing a $CaF_2$ tube, which was used for these experiments [36]. The cell is heated by three independent ovens. Ovens 1 and 2 control the temperature of the upper part of the cell, whereas oven 3 controls the temperature of the Cs reservoir and therefore the Cs vapour pressure within the cell. Selective reflection is performed on the inner side of the tube, while the other side, close to the YAG window, is kept at low temperatures to avoid the presence of Cs vapour in the gap between the tube and the window.

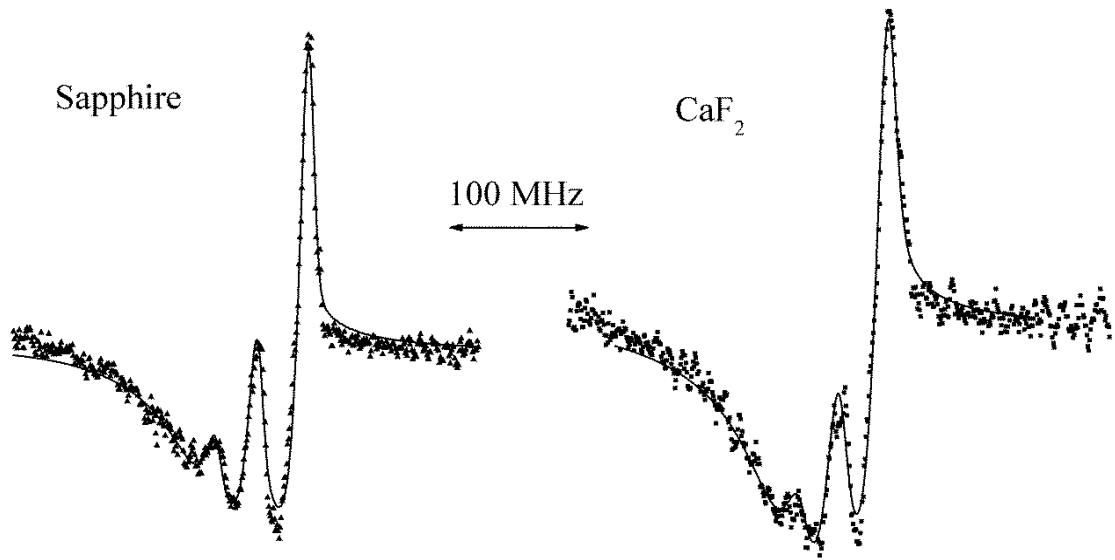

**Figure 4.** Selective reflection measurements performed on both sapphire and CaF$_2$ interfaces. The points represent experimentally measured spectra whereas the solid lines show the fits obtained using the theoretical model. In the case of sapphire we obtain the following values $C_3$=59 kHz µm$^3$, $\Gamma$=15.6 MHz and $\delta$=-6,3 MHz, whereas for CaF$_2$ $C_3$=66 kHz µm$^3$, $\Gamma$=23,5 MHz and $\delta$=-5,4 MHz.

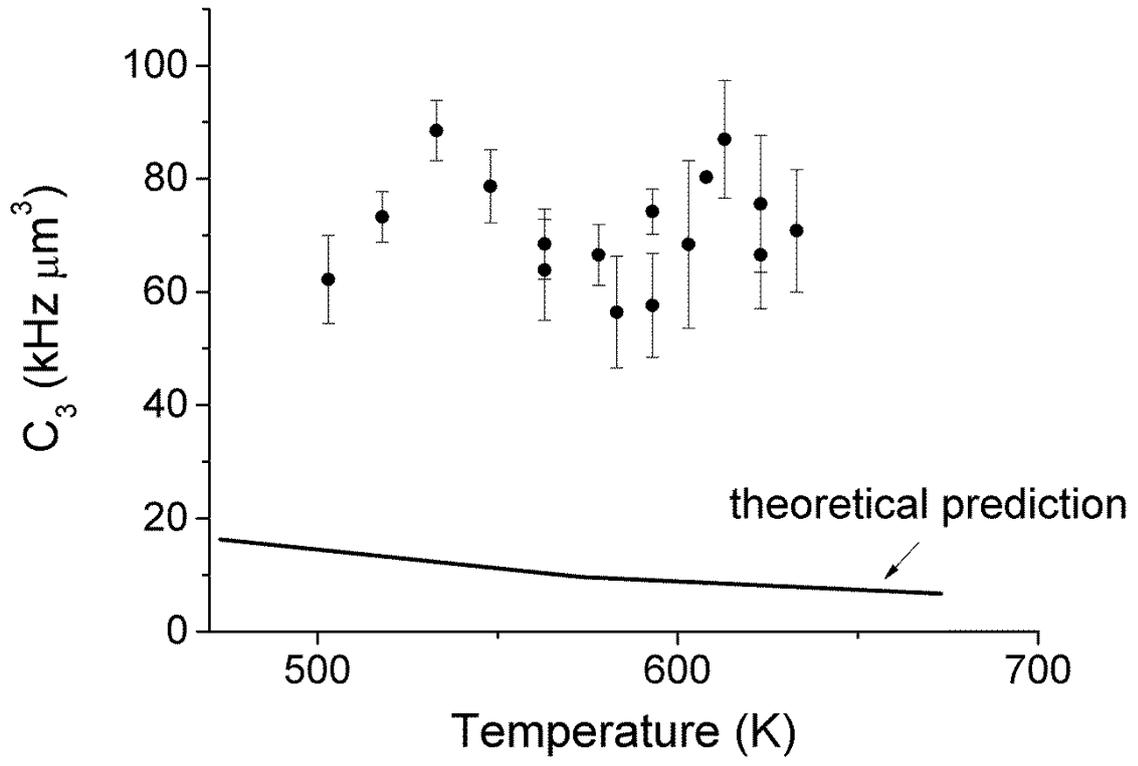

**Figure 5.** $C_3$ as a function of the $CaF_2$ window temperature.

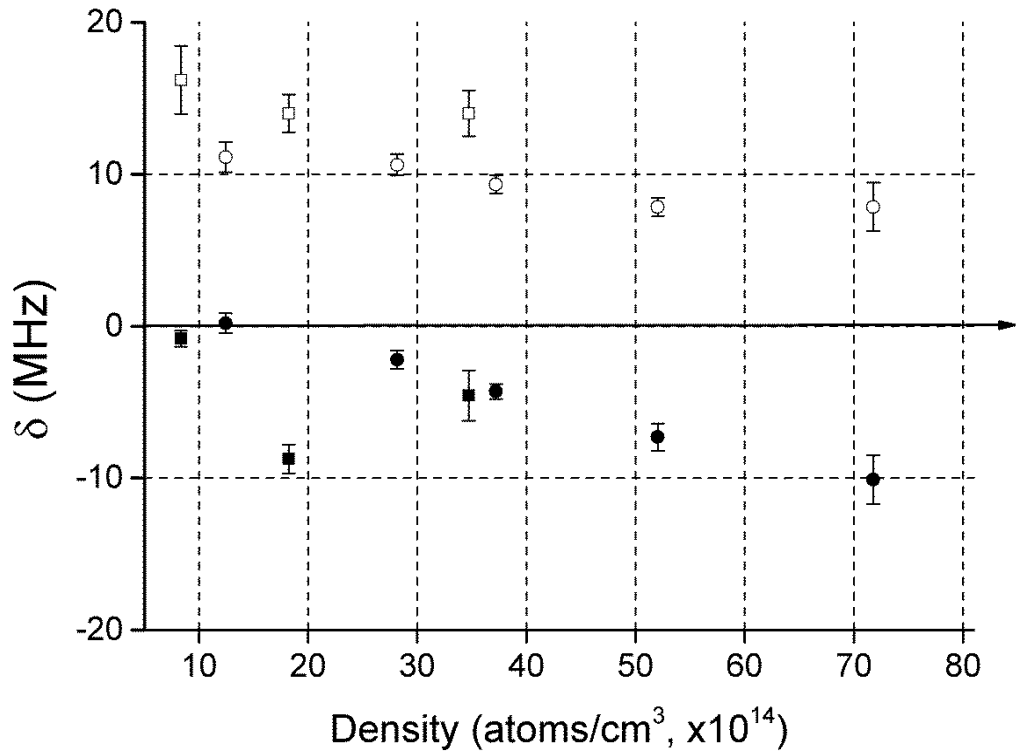

**Figure 6.** Values of the collisional shift as extracted from the fits of our experimental measurements. The solid circles (sapphire window) and squares ($CaF_2$ window) represent the collisional shift obtained when fitting with a $-C_3/z^3$ potential. The open circles and squares represent the results obtained for a $-C_2/z^2$ potential.

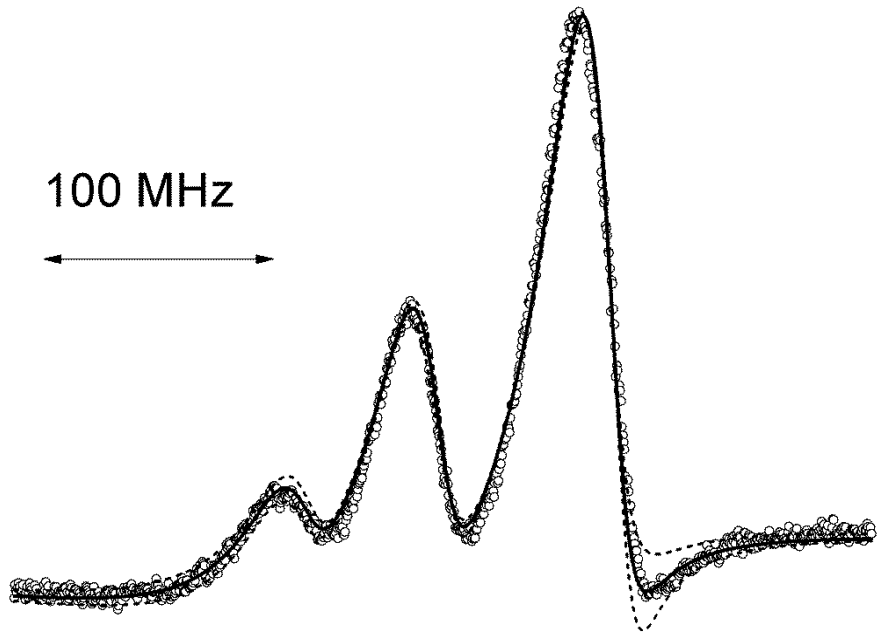

**Figure 7.** Selective reflection spectrum of the F=4 → F'=3,4,5 manifold of the $6P_{1/2}$→$7D_{3/2}$ transition. Experimental spectrum is shown with open dots. The sapphire window temperature was T=500 K. The best fit (solid line) is obtained for $C_3$=59 kHz µm³, Γ=17 MHz and δ=0. The dashed curves represent fits for which we impose a $C_3$ value different by ±15%.